\documentstyle[epsf]{elsart}
\begin{document}
\begin{frontmatter}
 \title{ Coexistence of monopoles and instantons for different 
         topological charge definitions and lattice actions }
\author[wien,bu]{M. Feurstein},
\author[wien]{H. Markum} and 
\author[wien,bu]{S. Thurner} 
\address[wien]{
 Institut f\"ur Kernphysik, TU Wien, Wiedner  Hauptstra\ss{}e 8-10, 
 A-1040 Vienna, Austria }
\address[bu]{College of Engineering, Boston University, 44 Cumminton St.,
             Boston, MA 02215, USA}

\begin{abstract}
We compute instanton sizes and study correlation functions between
instantons and monopoles in maximum abelian projection within $SU(2)$ 
lattice QCD at finite temperature. We compare several definitions of the 
topological charge, different lattice actions and 
methods of reducing quantum fluctuations. 
The average instanton size turns out to be $\sigma \approx 0.2$ fm. 
The correlation length between monopoles and instantons is 
$\zeta \approx 0.25$ fm and hardly affected by lattice artifacts 
as dislocations. 
We visualize several specific gauge field configurations
and show directly that there is an enhanced probability for finding
monopole loops in the vicinity of instantons. 
This feature is independent of the topological charge definition used.
\end{abstract}
\end{frontmatter}
\section{Introduction}
Classical gauge field configurations with non-trivial topology
are believed to play an essential role in the confinement mechanism.
In the scenario of the dual
superconductor abelian monopoles condense leading to confinement.
Large and interacting instantons could also produce confinement if they
form an instanton liquid. 
At first sight these two pictures are distinct and the interesting 
question arises, whether instantons and monopoles
are related to each other. 
Several groups investigated the  relation between monopoles and instantons 
for semi-classical configurations \cite{chernodub,suganuma}.
We presented first evidence that those correlations also exist in
realistic equilibrium configurations \cite{physrevd}. 

In this letter we compute instanton sizes from  auto-correlation functions 
of topological charge densities. We calculate correlation functions 
between topological charge densities and monopole currents. 
For the extraction of the topological charge we use 
field theoretical methods and the geometrical L\"uscher charge definition. 
In addition to the standard Wilson plaquette action we discuss 
an eight parameter fix-point action \cite{hasenfratz}.
From the auto-correlation functions of the topological charge 
instanton sizes can be estimated in smooth gauge-field configurations. 
Smoothing is done by cooling (for the Wilson action) and by 
means of blocking and reverse blocking (for the fix-point action) 
which is called constrained smoothing \cite{berlin96}.  
The correlation between monopoles and instantons is analyzed 
for the different topological charge definitions
and after suppression of dislocations. 
Correlation functions are obtained both in the confinement and 
deconfinement phase of pure QCD. 
For direct insight into the local geometry of topological activity we 
visualize specific configurations by tools of computer graphics. 
\section{Topological Charges and Monopoles}
There is no unique way to define a topological charge operator on the lattice.
The field theoretical methods are straightforward discretizations of
the continuum charge density:  
\begin{equation}
q(x)=\frac{g^{2}}{32\pi^{2}} 
\epsilon^{\mu\nu\rho\sigma}
\ \mbox{\rm Tr} \ \Big ( F_{\mu\nu}(x) F_{\rho\sigma}(x) \Big ) \ .
\end{equation}
For our purposes we  employ the plaquette and the hypercube method 
\cite{divecchia},
consisting of a sum of products of link variables along two 
perpendicular plaquettes or along a hypercube, respectively. 
The topological charge obtained has to be renormalized. A possibility 
to get rid of the renormalization constants is to use the method of 
cooling which reduces quantum fluctuations iteratively. Cooling, 
however, not only eliminates quantum fluctuations, but also small instantons
and  even large lattice instantons which die out if 
cooling is performed too intensely \cite{kronfeld}.
In particular we use  the ``Cabbibo-Marinari cooling method'' with a cooling
parameter of $\delta = 0.05$.

A way to overcome the problems associated with cooling is to use 
geometrical methods that interpolate the discrete set of link variables 
to the continuum in order to reconstruct the principal fibre bundle. 
We employ  the locally gauge invariant 
L\"uscher charge definition for $SU(2)$ \cite{schier_l}.
A drawback of the geometrical methods is that for the Wilson action  they 
are plagued by topological defects on the scale of the lattice spacing, 
dislocations. 
It it therefore necessary to smooth the gauge fields. 
To estimate the influence of dislocations on our results with the 
Wilson action, we use  a simplified 
fix-point action that suppresses dislocations \cite{hasenfratz}.
To measure the sizes of instantons it is still necessary to 
suppress quantum fluctuations.
To this end constrained smoothing based on renormalization group 
transformations was proposed as an alternative method to cooling 
which preserves long range physics and leaves instantons invariant 
\cite{berlin96}.

In order to investigate monopole currents we project $SU(2)$
onto its  abelian degrees of freedom, such that an abelian $U(1) $
theory remains \cite{thooft2}. This can be achieved by
various gauge fixing procedures. We employ the
so-called maximum abelian gauge which is most favorable for our
purposes.
In our simulations we subjected the configurations to 300 gauge fixing steps.
For the definition of the monopole currents $m(x,\mu)$ we use the
standard method \cite{SCH87}. After fixing the gauge the abelian parallel
transporters $u(x,\mu)$ are extracted and the color magnetic currents are
computed: 
\begin{equation}
m(x,\mu) = \frac{1}{2\pi} \sum_{\Box \ni \partial f(x+\hat\mu,\mu)}
\mbox{\rm arg } u(\Box) \ ,
\end{equation}
where $u(\Box)$ denotes a product of abelian links $u(x,\mu)$ around
a plaquette $\Box$ and $f(x+\hat\mu,\mu)$ is an elementary cube perpendicular
to the $\mu$ direction with origin $x+\hat\mu$.
From the monopole currents we define the local monopole density as
$ \rho(x) = \frac{1}{ 4V_{4}} \sum_{\mu} | m(x,\mu) | \ .  $
\section{ Correlation Functions }
Our simulations were performed on a $12^{3} \times 4$ lattice with
periodic boundary conditions using the Metropolis algorithm.
The observables were studied for the Wilson action both in the confinement
and the deconfinement phase at inverse gluon coupling
$\beta=4/g^{2}=2.25$ ($T/T_c=0.88$) and $2.4$ ($T/T_c=1.29$), respectively.
For the fix-point action  the inverse coupling ranged from 
$\beta=4/g^{2}=1.50$ ($T/T_c=0.83$) to $1.8$ ($T/T_c=1.73$). 
 For each run  we made 100  measurements, separated by 100 and 20 iterations 
for the Wilson action and  for the fix-point action, respectively. 

The normalized auto-correlation functions $\langle q(0) q(r) \rangle$ 
of the topological charge density are displayed in Fig.~1. 
In the case of the Wilson action they are presented 
for the hypercube definition (left) and for the L\"uscher method (middle) 
for 0, 5, 20 cooling steps in the confinement phase at $\beta=2.25$. 
Without cooling both auto-correlation functions are $\delta$-peaked
due to the dominance of  quantum fluctuations.
They become  broader with cooling reflecting the existence
of extended instantons. The auto-correlation function of the hypercube 
charge density is broader than that of the  L\"uscher charge density, 
because the hypercube
charge operator is more extended than L\"uscher's. 
The auto-correlation of the L\"uscher charge using 
a fix-point action is shown on the right-hand side of Fig.~1 in 
the confinement phase at $\beta=1.50$ 
before and after constrained smoothing. It is again
a $\delta$-function for the original configurations 
and broadens after performing the smoothing procedure 
because quantum fluctuations are drastically reduced \cite{hep_hase}. 
Here it is not necessary to finetune the cooling parameter and 
the number of cooling steps. 
\begin{figure*}[t]
\begin{center}
\begin{tabular}{ccc}
{\bf Wilson action} & {\bf Wilson action} 
& {\bf Fix-point action}   \vspace{-0.3cm} \\   
 {\small Hypercube charge}   &{\small L\"uscher charge}    
&{\small L\"uscher charge} \vspace{+0.5cm} \\ 
\hspace{-0.5cm}\epsfxsize=4.5cm\epsffile{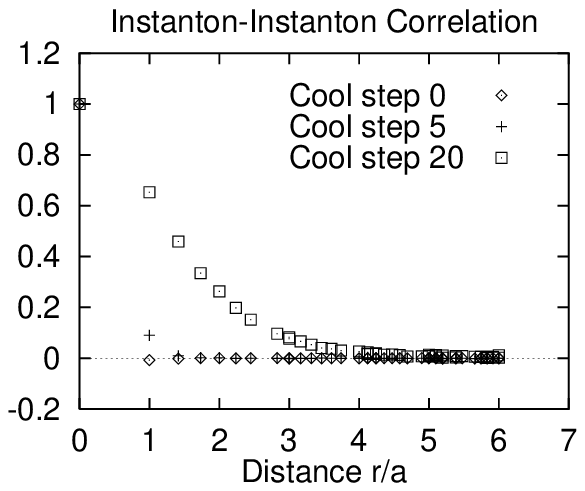} &
\hspace{-0.5cm}\epsfxsize=4.5cm\epsffile{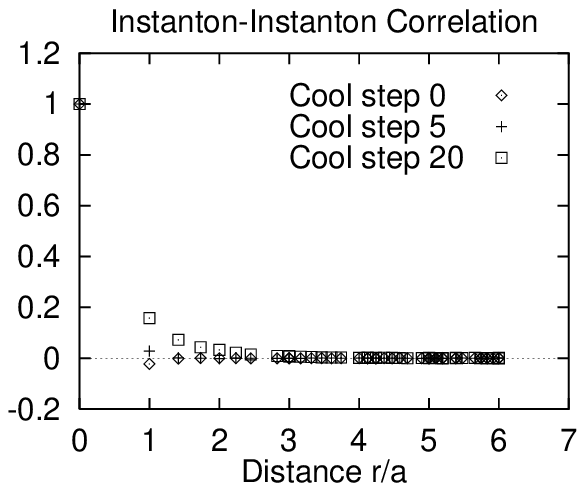} &  
\hspace{-0.5cm}\epsfxsize=4.5cm\epsffile{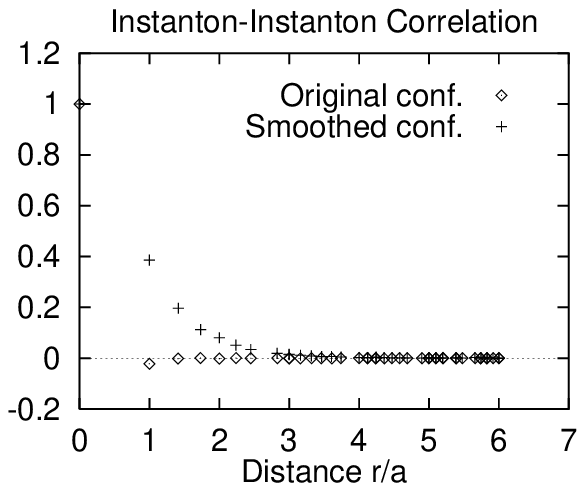} \\ \\
\end{tabular}
\end{center}
\vspace{-0.5cm}
\caption{
~Auto-correlation functions of the topological charge density
in the confinement phase using the hypercube definition (left) 
and the L\"uscher definition (middle) for 0, 5, 20 cooling steps 
for the Wilson action.  
Correlations for the L\"uscher charge definition before 
and after constrained smoothing for the fix-point action (right).
}
\end{figure*}

To obtain an average instanton size we fitted the 
auto-correlation functions of the L\"uscher charge 
to the convolution $f(x)=\int Q_\sigma(t)Q_\sigma(x-t)\, dt$
of the topological charge density  
$ Q_\sigma(x)= \frac{6}{\pi^2 \sigma^4}\,\, (\frac{\sigma^2}{x^2 + \sigma^2})^4$ 
of a single instanton with size $\sigma$. 
Such a fit is justified if instantons are dilute and  well separated 
which is the
case after 20 cooling sweeps or after constrained smoothing. 
The fitted instanton sizes $\sigma$  are displayed in physical units
for various $\beta$ values in Table~1. The errors are in the range of 15\%. 
For the Wilson action we used the 2-loop formula 
$a(\beta) = g(\beta)/\Lambda$ with $\Lambda=6.6$ MeV 
to obtain physical units.
For the fix-point action we extracted the lattice spacing $a$ at each
$\beta$ value from the string tension $(440  {\rm MeV})^2$ at $T$=0. 
Instantons show a tendency to become smaller on average 
with increasing temperature crossing the phase transition. 
However the results obtained with cooling seem less reliable due to the 
freedom in the cooling parameter. 
\begin{table}[b]
\label{instan}
\caption{Instanton sizes $\sigma$ for different values of $T/T_c$ 
computed from the auto-correlation functions of the  L\"uscher charge 
for the Wilson action after 20 cooling steps and for the fix-point action 
after constrained smoothing.}
\vspace{2mm}
\begin{center}
\begin{tabular}{|c|c|c|c|c|c|c|}
\hline
$T/T_c$ & 0.83 & 0.88& 0.92 & 1.09 & 1.29 & 1.73 \\
\hline
\hline
$\sigma$ [fm] (Wilson action) 
& ---  & 0.20 & --- & ---& 0.12& --- \\
$\sigma$ [fm] (Fix-point action)
&  0.31 & ---& 0.27& 0.21& --- & 0.10\\
\hline
\end{tabular}
\end{center}
\end{table}

As a measure for the local relation between abelian monopoles and instantons, 
we calculate~the~cor\-relation func\-tions $\langle |q(0)|\rho(r) \rangle$
between the absolute value of the topological charge density  
and the monopole density. 
They are displayed in Fig.~2 both in the confinement
and the deconfinement phase, after subtracting the 
cluster value and normalization. 
For the Wilson action the correlation
functions are computed employing the hypercube method (left) and the L\"uscher
method (middle) for several cooling steps. 
The shape of these correlations hardly changes under the influence 
of cooling and is essentially unaffected by the phase transition. 
Again the  correlation functions with the hypercube charge are somewhat broader 
than those with the L\"uscher charge due to the different extensions of the 
operators. 
\begin{figure*}[t]
\begin{center}
\begin{tabular}{ccc}
{\bf Wilson action} & {\bf Wilson action} 
& {\bf Fix-point action}   \vspace{-0.3cm} \\   
 {\small Hypercube charge}   &{\small L\"uscher charge}    
&{\small L\"uscher charge} \vspace{+0.5cm} \\ 
\hspace{-0.5cm}\epsfxsize=4.5cm\epsffile{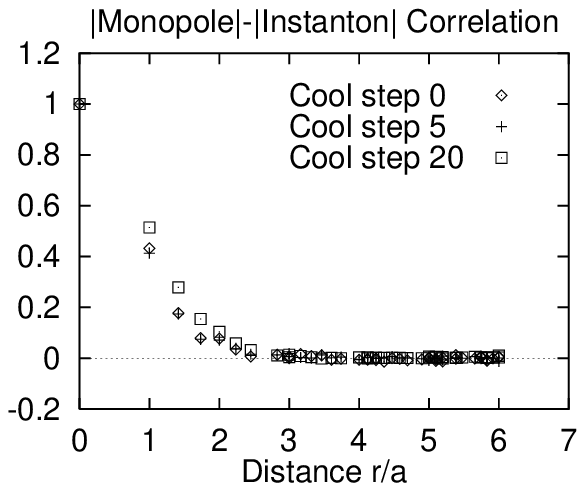} &
\hspace{-0.5cm}\epsfxsize=4.5cm\epsffile{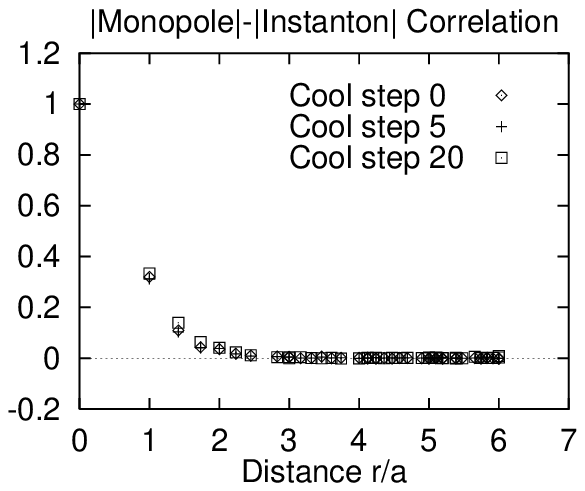} &   
\hspace{-0.5cm}\epsfxsize=4.5cm\epsffile{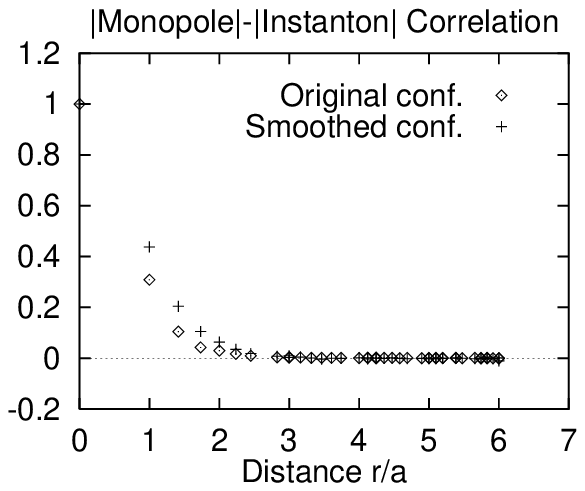} 
\vspace{-0.0cm}\\ 
\hspace{-0.5cm}\epsfxsize=4.5cm\epsffile{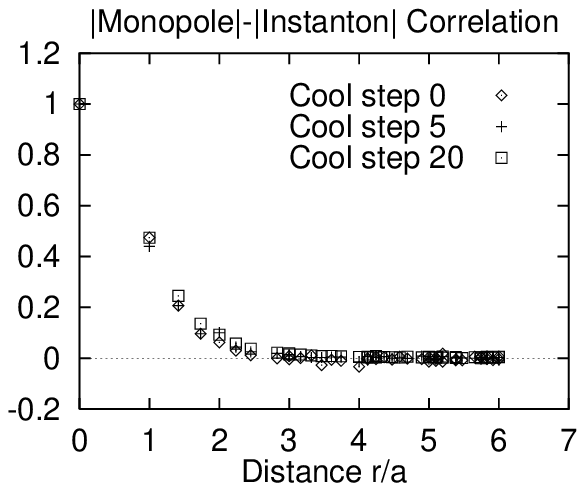} &
\hspace{-0.5cm}\epsfxsize=4.5cm\epsffile{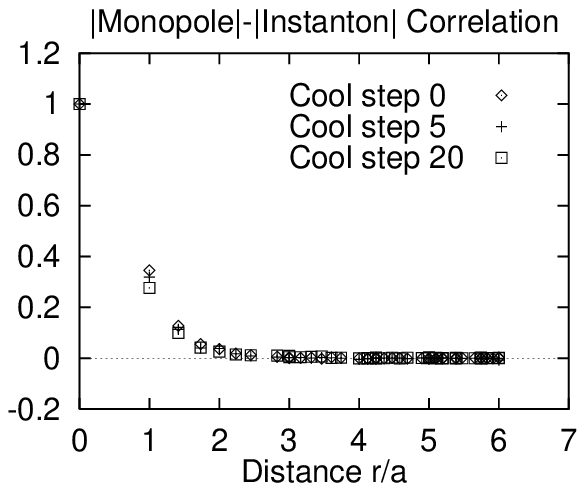} & 
\hspace{-0.5cm}\epsfxsize=4.5cm\epsffile{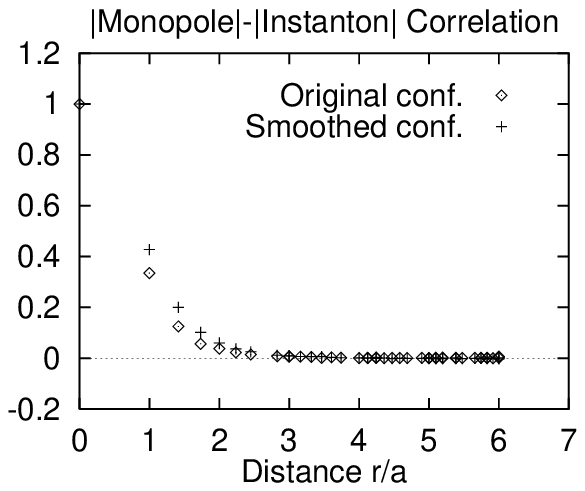}\\ \\
\end{tabular}
\end{center}
\vspace{-0.5cm}
\caption{
~Correlation functions between the monopole density and the absolute 
value of the topological charge density for the hypercube definition 
(left) and the L\"uscher definition (middle)  for 0, 5, 20 cooling 
steps for the Wilson action.
Correlations using the L\"uscher definition and the 
fix-point action before and after constrained smoothing (right). 
The correlations are computed in the confinement phase (above) 
and in the deconfinement phase (below).
}
\end{figure*}

L\"uscher's charge together with the Wilson action is known to suffer 
from dislocations which might have a non-trivial correlation with
monopoles \cite{schier_disl}. To get rid of dislocations that may spoil 
the physical results, we use the fix-point action. The correlation
functions of the L\"uscher charge distribution with the monopole density 
before and after constrained smoothing at $T/T_c=0.83$ and $T/T_c=1.09$ 
are depicted on the right-hand side of Fig.~2. 
They turn out to be similar to those from the 
Wilson action and become slightly wider after smoothing. 
Therefore dislocations are not decisive for 
the non-trivial correlation found for the Wilson action. 
For all cases (Wilson action, fix-point action, cooling, constrained smoothing,
confinement and deconfinement phase) the correlation lengths in lattice units
are rather similar. 
The correlation length has a tendency to increase with temperature 
and was found to cover a range of $\zeta=0.15-0.35$ fm. 
 
It has been reported that the ratio 
$R= \frac{\rho_s}{3\times\rho_t}$ of space-like to time-like 
monopole densities decreases across the deconfinement phase transition
and that it might serve as a reasonable order parameter \cite{suz}.
We observe that the same quantity also
decreases as a function of cooling which is displayed in Table 2. 
The drastic decrease yields some doubt on the quality of this
quantity as an order parameter. 
For example after 20 cooling sweeps the
string tension of $SU(2)$ at $\beta =2.25$ is still present, even
though $55\%$ of the monopole currents are static. However at $\beta =2.4$
in the uncooled configurations only $27\%$ of the monopoles are static
but the string tension vanishes completely.
\begin{table}[t]
\label{moncool}
\caption{Ratio of spatial to time-like monopole density
in the confinement and deconfinement for various cooling steps. }
\vspace{2mm}
\begin{center}
\begin{tabular}{| c | c | c|}
\hline
{\small Cool step} 
             & $\beta$ = 2.25 &  $\beta$ = 2.40          \\
\hline
\hline
0  &  0.996 & 0.898  \\
5  &  0.975 & 0.571  \\
20 &  0.268 & 0.037 \\
\hline
\end{tabular}
\end{center}
\vspace{2mm}
\end{table}
\section{Visualization} 
We visualize the relation between instantons and monopoles 
by directly displaying clusters of topological charge and 
by drawing monopole loops in fixed time slices of specific configurations.
For any value of the topological charge density $q(x) > 0.01$ a light dot 
and for $q(x) < 0.01$ a dark dot is plotted. 
The lines represent the monopole loops.
  
In Fig. 3 we compare the results of the field theoretical methods and the 
L\"uscher method for the Wilson action on a single equilibrium
gauge field after 20 cooling sweeps. From left to right 
the plaquette, hypercube, and the L\"uscher charge distributions are plotted.
The positions of the clusters of topological charge are the same for
all methods. The points represent instantons or anti-instantons.
In this particular configuration a monopole is found to
wrap around the torus.
\begin{figure}[t]
\begin{center}
\begin{tabular}{ccc}
Plaquette definition \vspace{-0.0cm} & 
Hypercube definition \vspace{-0.0cm} & 
L\"uscher definition \vspace{-0.0cm}\\ 
\epsfxsize=4.0cm\epsffile{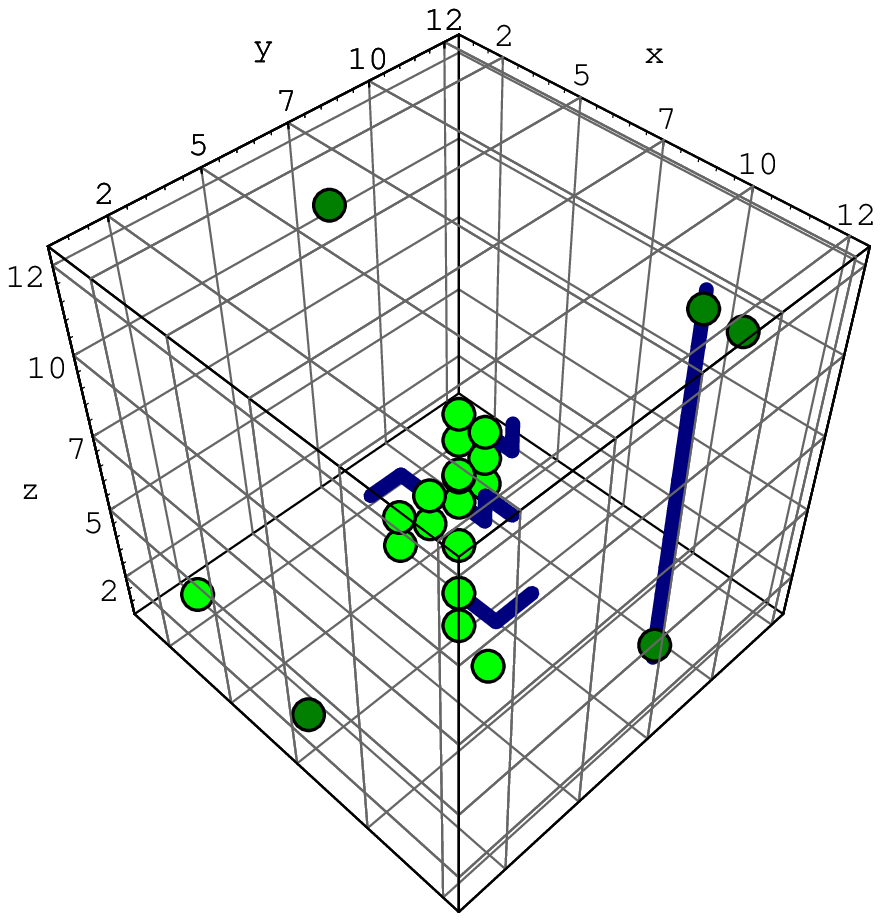} &
\epsfxsize=4.0cm\epsffile{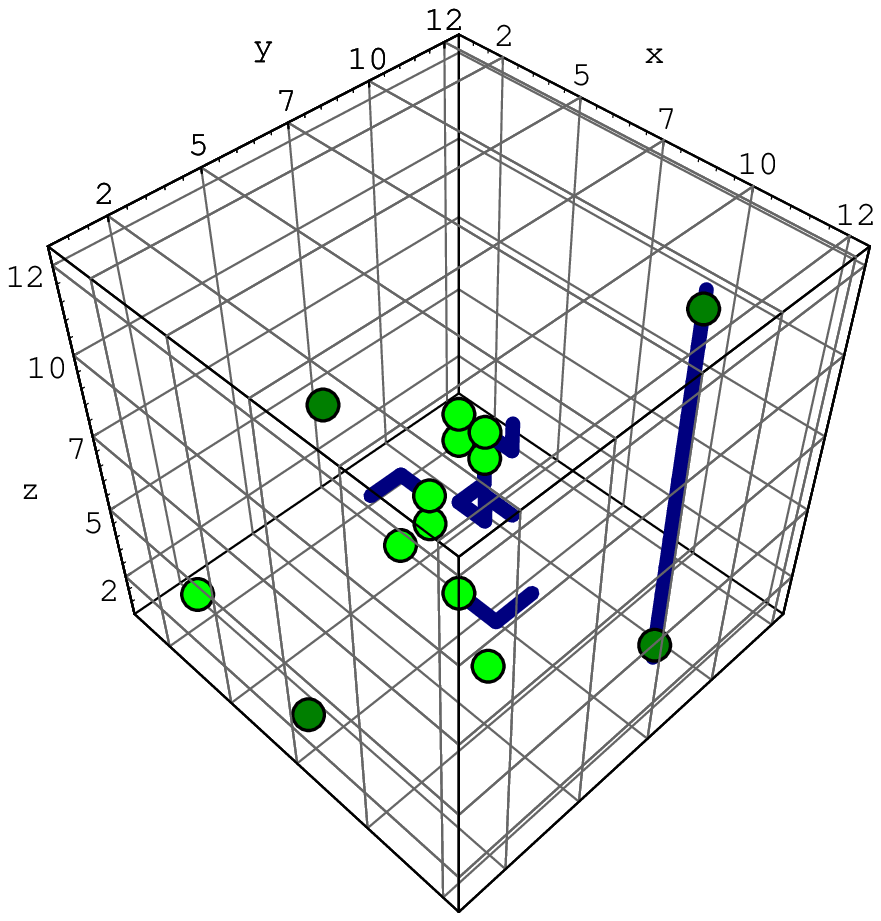} &
\epsfxsize=4.0cm\epsffile{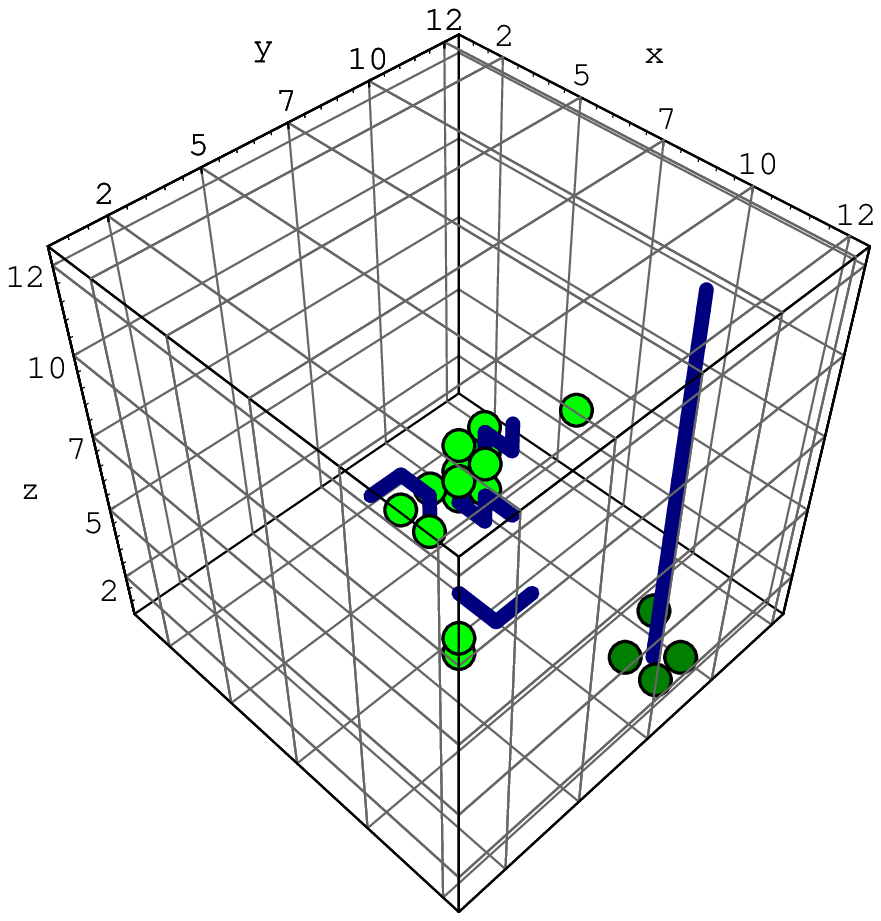} \\
\end{tabular}
\end{center}
\caption{
~Different definitions of the topological  charge
for a specific gluon configuration from Wilson action after $20$
cooling sweeps. The instantons reside at the same places for all definitions
and are surrounded by monopoles.
In this configuration a monopole loop wraps around the torus.
}
\end{figure}
 
Fig. 4 presents a cooling history of a time slice of a gluon field.
The topological charge using the L\"uscher definition 
with the Wilson action is displayed for cooling steps 0, 15, 20 and 25.
Without cooling the topological charge distribution
cannot be identified with instantons due to quantum fluctuations. 
After  15-20 cooling steps one can assign instantons to
clusters of topological charge. At cooling  sweep 20 an
instanton and an anti-instanton emerge. From cooling steps 35-40 they begin 
to approach each other and  annihilate several  steps later (not shown).
Monopole loops also thin out with cooling, but they survive in the
presence of instantons. In general, there is 
an enhanced probability that monopole loops are present in the vicinity of
instantons in all gauge field configurations.
\begin{figure*}[t]
\begin{center}
\begin{tabular}{cc}
\vspace{-0.5cm}\\
0 cooling sweeps & 15 cooling sweeps  \vspace{-0.2cm}\\
\epsfxsize=5.0cm\epsffile{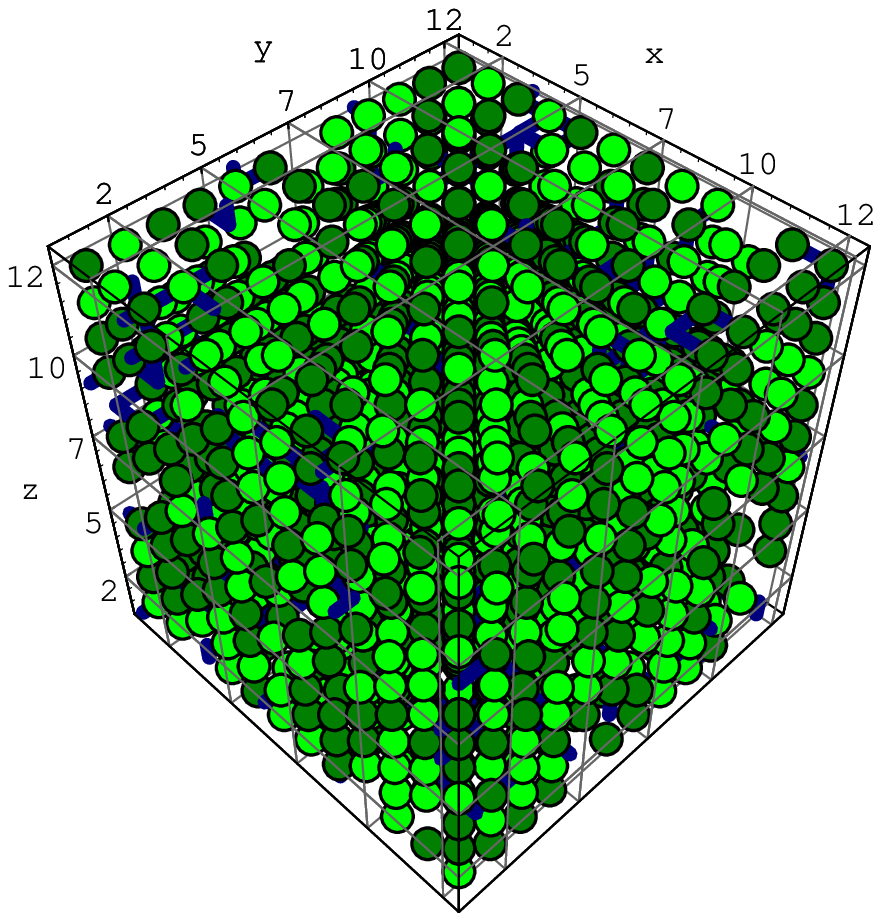}
&
\epsfxsize=5.0cm\epsffile{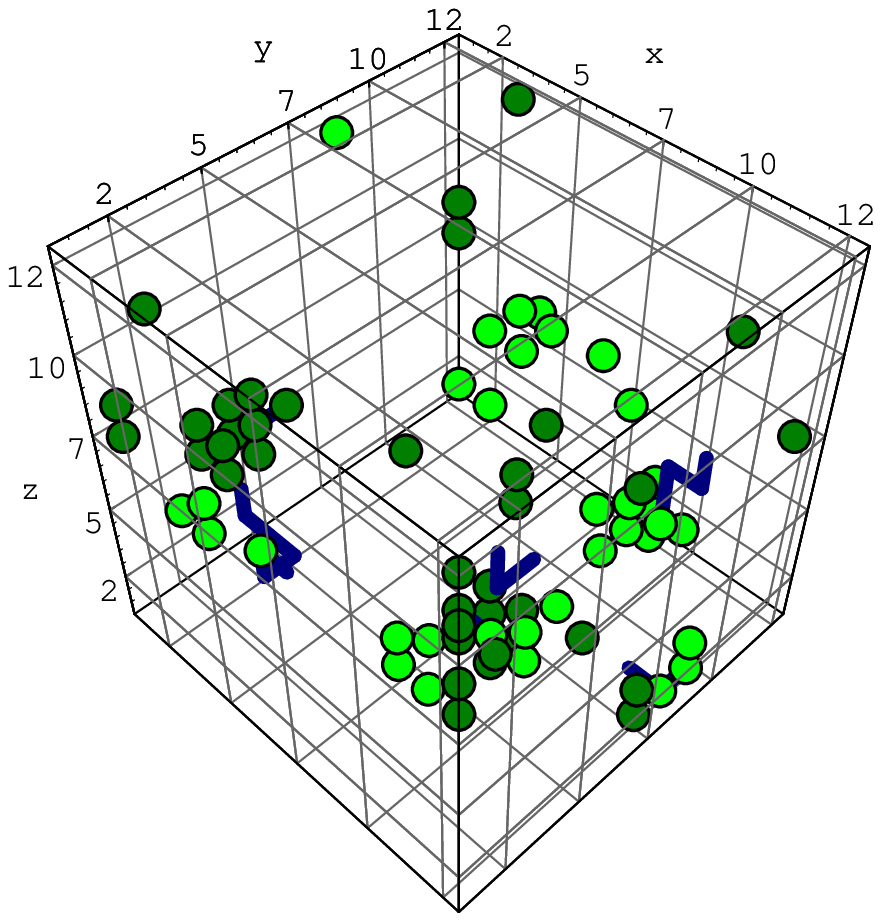}
\vspace{0.6cm}\\
20 cooling sweeps & 25 cooling sweeps \vspace{-0.2cm}\\
\epsfxsize=5.0cm\epsffile{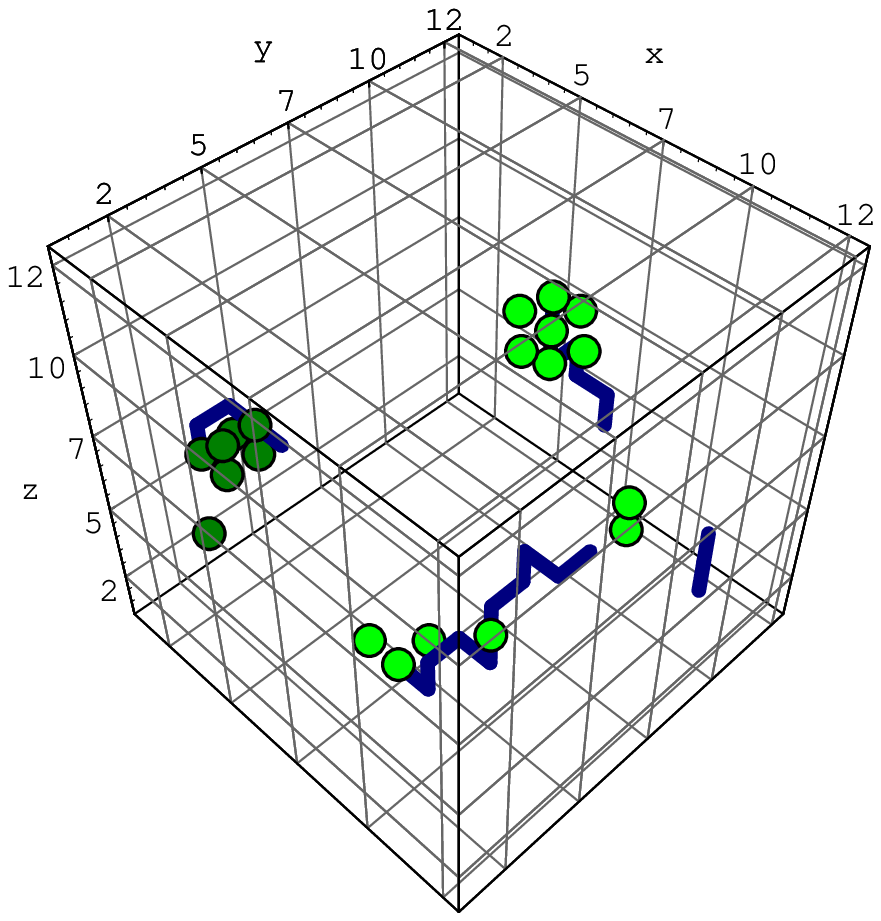}
&
\epsfxsize=5.0cm\epsffile{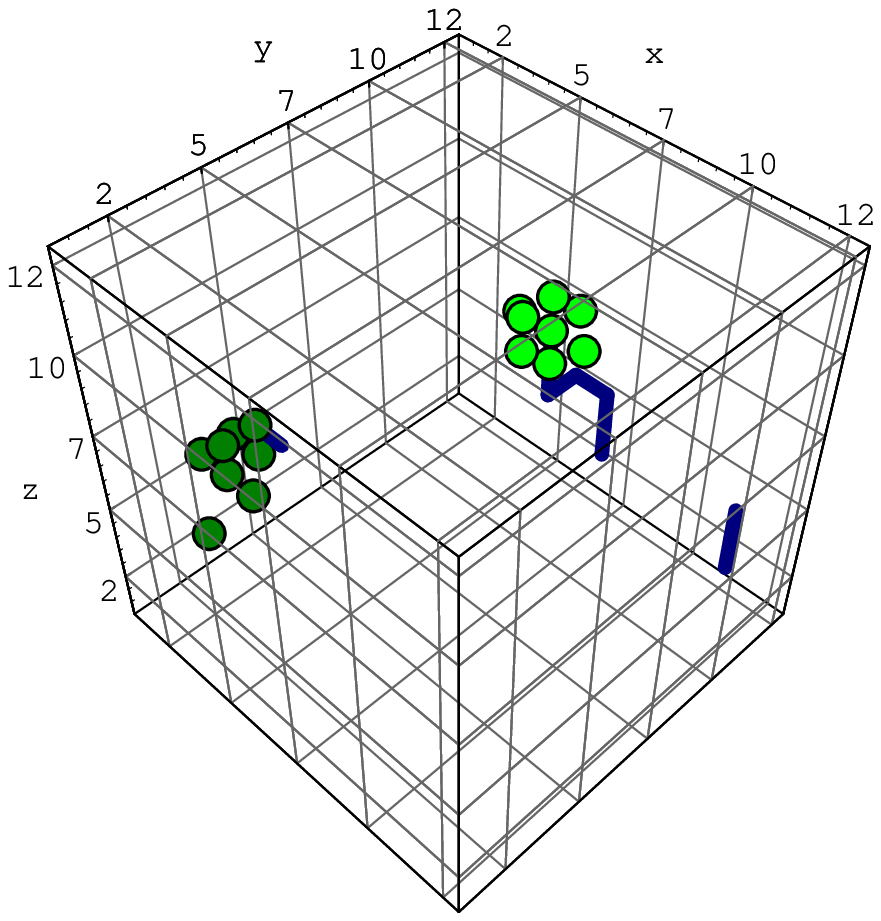} \vspace{0.6cm}\\
\vspace{0.3cm}\\
\end{tabular}
\end{center}
\vspace{-1.0cm}
\caption{
~Cooling history for a specific gauge field configuration from Wilson action 
at a fixed time slice. The dots represent the topological charge distribution 
in the L\"uscher definition with $|q(x)|>0.01$. 
Positive (negative) charges are plotted with a light (dark) dot. 
The monopole loops correspond to lines. With cooling
instantons evolve from noise accompanied by monopole loops
in almost all cases. Note that time-like monopoles cannot be seen in
these plots.
}
\end{figure*}
\section{Conclusion} 
We computed average sizes of instantons and correlation functions between
instantons and monopoles for different topological charge definitions,
for different lattice actions, for cooled and constrained smoothed 
configurations.  
The auto-correlation functions between the topological charge density 
yield consistent results of instanton sizes for the Wilson and fix-point 
action. In the confinement phase the instanton size ranges from 
$\sigma=0.2-0.3$ fm whereas in the deconfinement phase $\sigma=0.1-0.2$ fm. 
The correlation functions between abelian monopoles and instantons are 
very similar for the geometrical and the field theoretical definition. 
They are hardly affected by cooling (Wilson action) or by constrained smoothing
(fix-point action) and are qualitatively the same even across the 
deconfinement phase transition.   
The correlation length was found $\zeta=0.15-0.25$ fm in the confinement 
and $\zeta=0.25-0.35$ fm in the deconfinement phase. 

We further calculated the local values of topological charges and monopole 
currents and directly displayed them with the help of computer graphics. 
After a few cooling sweeps one 
observes clearly that instantons are accompanied by monopole loops. 
This correlation occurs on all (semi-classical) gauge field configurations 
considered. In a cooling history we demonstrated how instantons evolve 
from fluctuating gauge fields and how they are surrounded by monopoles.
Combining the above finding that the correlations are rather insensitive under
cooling or smoothing together with that of the 3D images, we conclude that the
topological charge goes hand in hand with monopoles also in the original 
gauge field configurations.
The demonstration of our simulation together with analytical investigations 
\cite{chernodub,seiberg} might present a first indication of a deep relation 
between the topological structure of compact abelian and non-abelian 
gauge field theories. Since in abelian theories monopoles are responsible 
for confinement, and if such a relationship existed, this could be accepted 
as a topological proof of quark confinement. 
\section{Acknowledgments}
We appreciate the cooperation with E.-M.~Ilgenfritz and M.~M\"uller-Preu\ss ker 
at Humboldt University Berlin where part of the work was done. 
We thank G.~Schierholz who provided us with a routine to compute
the L\"uscher charge and W.~Sakuler for continued interest. 
This work was partially supported by FWF under Contract No.~P11456-PHY. 


\end{document}